\documentclass[11pt]{article}
\usepackage{float}
\usepackage[symbol]{footmisc}
\usepackage[superscript,sort]{cite}
\usepackage{array}
\usepackage[normalem]{ulem} 
\usepackage[colorlinks=true,linkcolor=black,citecolor=black,urlcolor=black,bookmarks=true,breaklinks=true]{hyperref}

\usepackage{mathtools}
\usepackage{amsthm,amscd,amsxtra,amsfonts,amsmath,amssymb,multirow}
\usepackage{wrapfig}
\usepackage[footnotesize]{caption}
\usepackage{subcaption}
\usepackage[tiny,compact]{titlesec}
\usepackage{booktabs}
\usepackage{siunitx}
\usepackage{threeparttable} 
\usepackage{tabularx}
\usepackage{helvet}

\usepackage{xcolor,colortbl}
\usepackage{tikz}
\usetikzlibrary{shapes,arrows}
\usepackage[T1]{fontenc}
\usepackage[linesnumbered,ruled]{algorithm2e} 
\titlespacing*{\section}{0pt}{*0}{*0}
\titlespacing*{\subsection}{0pt}{*0}{*0}
\titlespacing*{\subsubsection}{0pt}{*0}{*0}
\titlespacing{\paragraph}{0pt}{*0}{*1}

\definecolor{MyPurple}{rgb}{1,0,1}

\newcommand{\barray}{\begin{array}{ll}}
\newcommand{\earray}{\end{array}}

\usepackage{mathtools}
\begin{document}
\pagenumbering{roman}

\clearpage \pagebreak \setcounter{page}{1}
\renewcommand{\thepage}{{\arabic{page}}}

\title{Review of Quantitative Systems Pharmacological Modeling in Thrombosis
}

\author{Limei Cheng$^1$ \footnote{Corresponds to Limei Cheng, Email: Limei.Cheng@bms.com.},
 Guo-Wei Wei$^{2}$ and Tarek Leil$^1$ 
	\\
	$^1$ Clinical Pharmacology and Pharmacometrics,
Bristol-Myers Squibb, NJ 08540, USA\\
	$^2$ Department of Mathematics,
	Michigan State University, MI 48824, USA.
}

\date{\today}
\maketitle

\begin{abstract}
  Hemostasis and thrombosis are often thought as two sides of the same clotting mechanism whereas hemostasis is a natural protective mechanism to prevent bleeding and thrombosis is a blood clot abnormally formulated inside a blood vessel, blocking the normal blood flow. The evidence to date suggests that at least arterial thrombosis results from the same critical pathways of hemostasis. Analysis of these complex processes and pathways using quantitative systems pharmacological model-based approach can facilitate the delineation of the causal pathways that lead to the emergence of thrombosis. In this paper, we provide an overview of the main molecular and physiological mechanisms associated with hemostasis and thrombosis, and review the models and quantitative system pharmacological modeling approaches that are relevant in characterizing the interplay among the multiple factors and pathways of thrombosis. An emphasis is given to computational models for drug development. Future trends are discussed. 
\end{abstract}
\maketitle



\section{Introduction}
 Hemostatic response to blood loss is a protective mechanism in all mammals to stop bleeding after an injury \cite{p1,p2,p3,p4}. The mechanisms of hemostasis involve series of biochemical cascades that are largely mediated by platelets. These systems can also be activated in a pathophysiological scenario by vascular malfunction, leading to obstructive blood clotting, namely thrombosis.  In thrombosis obstruction of blood flow through healthy veins and/or arteries can lead to a number of pathological situations.  Obstruction of arteries can cause serious ischemic injuries such as myocardial infarction, pulmonary embolism, and ischemic stroke.  Blockage of veins is typically less serious in the acute setting, but can cause venous thromboembolism (VTE) which can have long term consequences if untreated.  Many hemorrhagic and thrombotic complications are associated with chronic physiological conditions such as inflammation, infection, cancer, heart failure, obesity, and atherosclerosis. The high rates of thrombosis in these diseases mean that there is still a high unmet medical need in these populations, despite the numerous therapeutic options available to inhibit different mechanisms of thrombosis.  The long history of non-clinical and clinical research in thrombosis has led to a fairly comprehensive understanding of the biochemistry, molecular biology, systems biology, cellular mechanics, pathology, and physiology of hemostasis and thrombosis.  However, much of the quantitative and mechanistic understanding of thrombosis still remains elusive \cite{p3,p5}.  A more thorough quantitative and mechanistic characterization of hemostasis and thrombosis can help to uncover new mechanisms that can address the unmet medical need that still exists.

\subsection{Hemostasis and Thrombosis}

Hemostasis is the normal response to blood leaking. Initially, after injury, vascular smooth muscle cells produce vasoconstriction, then platelets start to adhere and aggregate on the surface of endothelium cells, forming platelet plug gradually, which is referred as primary hemostasis \cite{p1}.  Meanwhile, the clotting factors are activated through coagulation cascades to promote the formation of fibrin. Then, the produced fibrin mesh helps to hold the platelet plug in place to prevent blood leakage through the injury. The resultant plug,``thrombu'' or ``clot'', contains some red and white blood cells and therefore is harder than the primary hemostasis plug. This is referred as secondary hemostasis \cite{p2}. Although hemostasis is a positive and protective mechanism for wound healing, a similar clot can grow inside a blood-leak-free vessel for a variety of pathophysiological reasons, eventually resulting in thrombosis. Blood clot formation is stimulated by platelet aggregation and fibrin formation, involving numerous proteins and co-factors interacting with each other in addition to blood and vascular cells. Among them, platelet, fibrinogen, fibrin, and thrombin are the major players. Platelets are nucleus-free cell-like components of blood whose function is to react to bleeding from a blood vessel injury by clumping, thereby initiating the formation of a blood clot. Platelets function through cell adhesion, cell deformation, muscle-like platelet contraction, and fluid-gel transformation in addition to both intra-cellular and extra-cellular chemical reaction pathways \cite{p3}. The entire process is also complicated by the interplay between signal transduction pathways, molecular transport, fluid mechanic and cellular mechanics, for which there are little mechanistic experimental data \cite{p4}. Fibrin is an important protein in clot formation, as it forms a gel-like three-dimensional mesh atop the platelet plug, thereby completing the formation of the physical clot \cite{p4}. Fibrin generation is controlled by an intricate blood coagulation network, a cascade of proteolytic reactions involving a large number of biological mediators and balancing feedback loops \cite{p3}. The interactions between blood coagulation and platelet-dependent hemostasis are particularly sophisticated. For example, thrombin, as the main enzyme of hemostasis, not only catalyzes the fibrinogen-fibrin conversion, is also a major activator for procoagulant factors and protein C \cite{p3}. In fact, the spatiotemporal regulation aspect of this network is essentially unknown \cite{p5} although its biochemistry is well-studied. This unknown spatiotemporal regulation directly contributes to the difficulties in predicting the time and location of thrombosis. Typically, the formation of a thrombus is followed by termination of its spatial propagation, resulting in a localized clot that permits continued blood circulation within the vessel, and around the clot \cite{p5,p6}. However, uncontrolled thrombus generation gives rise to blockage of the blood vessel and downstream ischemic complications \cite{p3}. The chemical and biological mechanisms behind the spatiotemporal regulation of thrombus formation is an important area of active research \cite{p6,p7,p8,p9,p9,p10,p11}. Unfortunately, the critical pathways in hemostasis and arterial thrombosis are shared, which is the major reason that most anti-thrombotic drugs have serious bleeding side effects \cite{p7}. Although there are molecular or regulatory processes that are believed to be confined to either hemostasis or thrombosis, they are yet to be translated into safer and approved therapies.

\begin{figure}[h]
    \centering
    \includegraphics[width=1\textwidth]{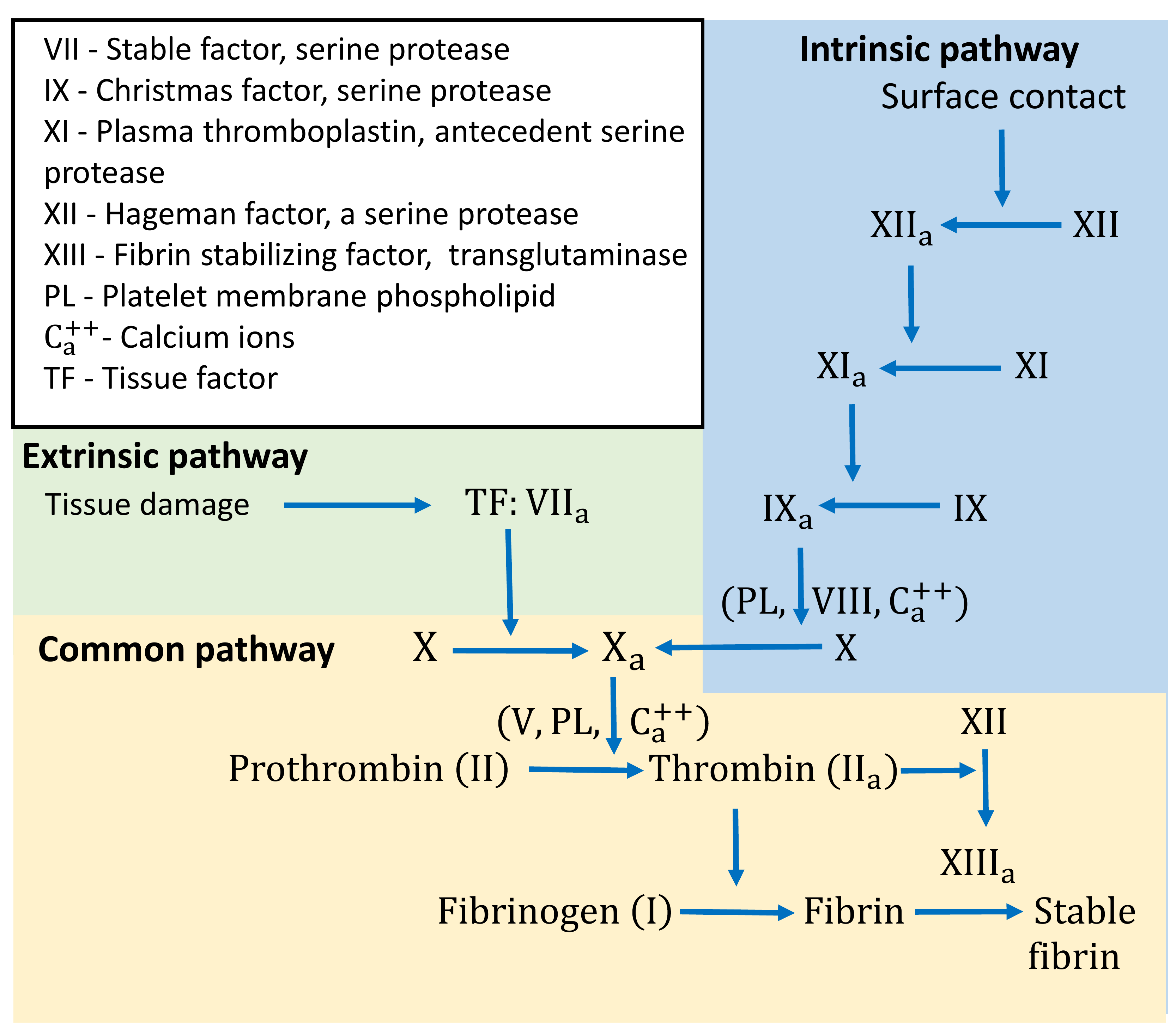}
    \caption{Schematic illustration of blood coagulation pathways \cite{p23}.  Both intrinsic pathway and extrinsic pathway lead to the final common pathway.}
    \label{fig:1}
\end{figure}
\subsection{Platelet Activation} 

The platelet thrombus cannot be formed without platelet activation \cite{p1}. One of the key initiation events for thrombus/clot formation is the interaction between platelets and subendothelial matrix proteins (von Willebrand factor (VWF), collagens types I, III and VI). The interactions of these proteins with platelet surface receptors, such as glycoprotein IIb/IIIa and VWF receptors, lead to several intracellular signaling cascades \cite{p12,p13,p14,p15}. Specifically, the binding of collagen to glycoprotein VI triggers a signaling cascade that results in the activation of integrins, which mediate the tight binding of platelets to the extracellular matrix. In such reactive network processes, platelets are transformed into several activated states associated with a series of responses, including adhesion and aggregation \cite{p1}. Activated platelets are known to respond with the exocytosis of the dense granules and alpha granules, including ADP, serotonin, platelet-activating factor (PAF), VWF, and platelet factor 4 (PF4), which, in turn, activate additional platelets. Moreover, the granules' contents activate a Gq-linked protein receptor cascade, resulting in increased calcium concentration in the platelets' cytosol \cite{p7}. Calcium signaling modulated microtubule/actin filament reorganization in the presence of ATP gives rise to platelet morphological changes required to adapt to the fluid dynamic and structural conditions within the blood vessel and thrombus \cite{p12}. Calcium also activates protein kinase C, which, in turn, activates the membrane enzyme phospholipase A2 (PLA2). PLA2 then modifies the integrin membrane glycoprotein IIb/IIIa, increasing its affinity to bind fibrinogen. PLA2 also promotes the formation of thromboxane A2 (TXA2), which stimulates the activation of new platelets. Activated platelets also cause the externalization of phosphatidylserine (PS) to orient coagulation proteases, specifically tissue factor (TF) and Factor VII which facilitate further proteolysis, activation of Factor X, and ultimately generating thrombin from prothrombin.  Details of the coagulation cascade are discussed in the next section.
\begin{figure}[h]
    \centering
    \includegraphics[width=1\textwidth]{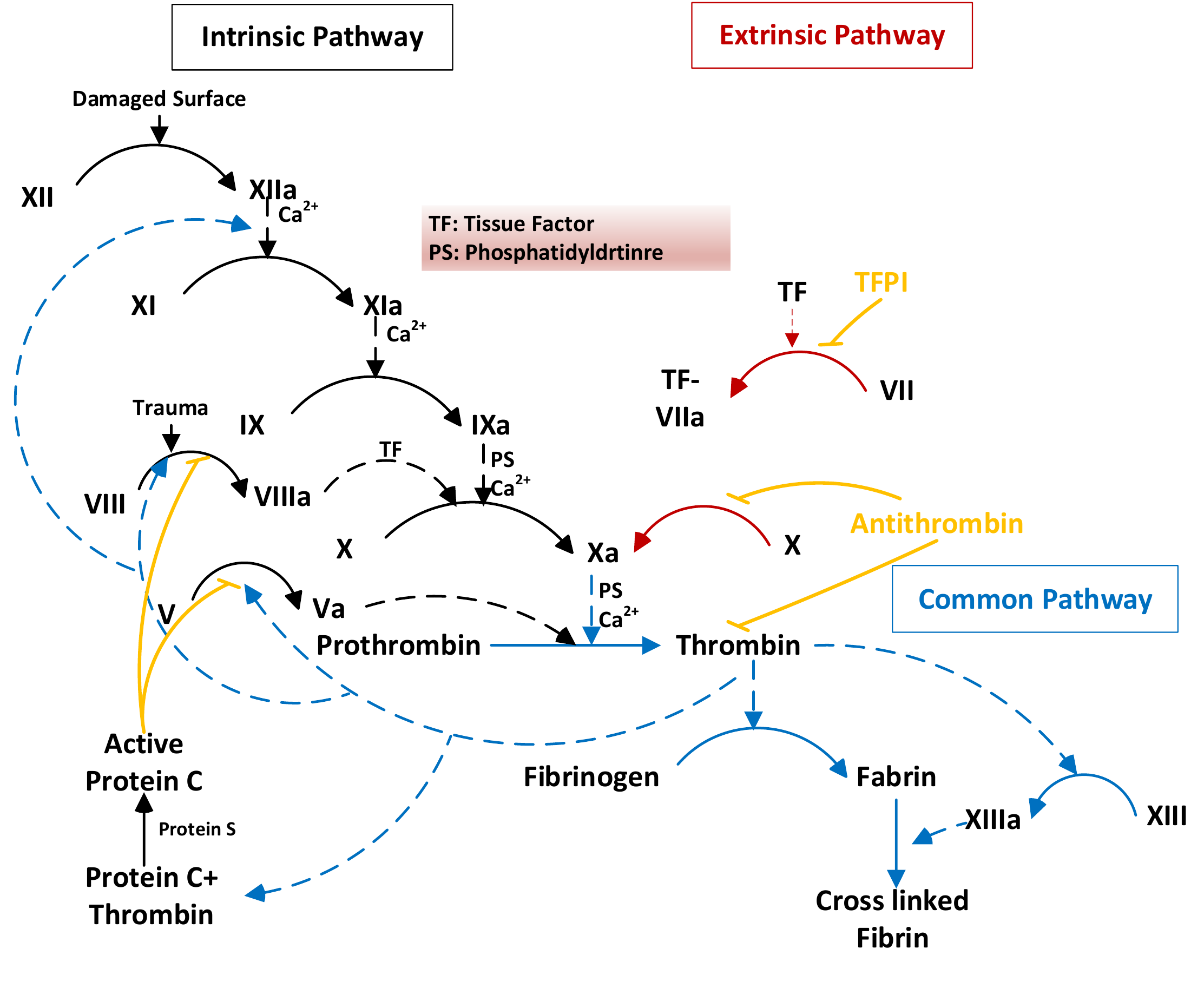}
    \caption{The systems biology of coagulation pathway networks showing arrows for negative and positive feedback. }
    \label{fig:2}
\end{figure}
\subsection{Coagulation Cascade} 
The blood coagulation network as depicted in Fig. \ref{fig:1} was discovered decades ago and the associated biochemistry is well-known \cite{p16}. The coagulation cascade of secondary hemostasis has two initial pathways, often termed extrinsic and intrinsic pathways; both of which lead to activation of Factor X and a final common pathway for fibrin formation \cite{p17}. The extrinsic pathway (also known as the tissue factor pathway) is the primary pathway for the initiation of blood coagulation by activating VII and IX, and then X. In contrast, the intrinsic pathway (also known as the contact activation pathway) activates sequentially Factors XII, XI, IX, and X. The intrinsic pathway is positively regulated by thrombin, which promotes the activations of V, VIII, and XI, leading to a positive feedback loop that can result in massive production of thrombin. Thrombin enzymatically converts fibrinogen (Factor I) to fibrin (Ia) and activates XIII to XIIIa, which stabilizes the fibrin mesh by cross-links. In fact, thrombin also binds to cell surface protein thrombomodulin, which activates protein C at the presence of protein S \cite{p16}. Then the activated protein C degrades Va and VIIIa, leading to a negative feedback loop as shown in Fig. \ref{fig:2}.

\subsection{Cofactors} 
The coagulation cascade cannot function without a number of cofactors \cite{p14,p15,p16}. Phospholipid, a platelet membrane constituent, is required for the tenase and prothrombinase complexes to function, which is crucial to Factor Xa and thrombin (Factor IIa) shown in Fig. 1. Another cofactor calcium is required in calcium signaling pathways and in other parts of the coagulation cascade, such as mediating the binding of the tenase and prothrombinase complexes to the platelet membrane phospholipid surfaces via terminal gamma-carboxylated glutamic acid (Gla) residues on Factor Xa and Factor IXa. Some roles of calcium are shown in Figs. \ref{fig:1} and \ref{fig:3}. One of the most important cofactors in the coagulation cascade is vitamin K. Vitamin K is essential in adding a carboxyl group to glutamic acid residues on Factors II, VII, IX and X, and Protein S, Protein C and Protein Z. It is also involved in the carboxylation of certain glutamate residues in proteins to form gamma-carboxyglutamate residues, which are usually involved in binding calcium. Oxidized vitamin K is reactivated by vitamin K epoxide reductase (VKORC), which is a pharmacological target of many anticoagulant drugs, including Warfarin and related coumarin anticoagulants such as Acenocoumarol, Phenprocoumon, and Dicumarol.  Vitamin K is used for the treatment of bleeding induced by Warfarin overdose.

\subsection{Regulators and fibrinolysis} 
In normal conditions, there are five regulators that keep platelet activation and the coagulation cascade in hemostasis to avoid the thrombotic tendency, as shown in Fig. 2. Protein C inhibits Factors VIIIa and Va and together with thrombin binding to a cell surface protein thrombomodulin to activate Protein C. At the presence of Protein S and phospholipid, the activated Protein C degrades Factors Va and VIIIa. Antithrombin, produced by the liver, is a small protein that inactivates several coagulation enzymes, including thrombin \cite{p17}, Factors IXa, Xa, XIa, and XIIa. Antithrombin forms a complex with a serine protease in which the active site of the protease enzyme is inaccessible to its typical substrate. Its activity is increased many folds by the anticoagulant drug Heparin, which enhances the binding of antithrombin to thrombin, Factors IXa, Xa, XIa, and XIIa. Tissue factor pathway inhibitor (TFPI), a single-chain polypeptide, limits the primary pathway of the action of tissue factor (TF), by inhibiting the activation of Factor VII, as shown in Fig. 2. Prostacyclin (PGI2) inhibits platelet activation. As shown in Fig. 3, it is released by endothelium and activates platelet prostaglandin receptors. This, in turn, activates adenylyl cyclase (ADCY3), which synthesizes the secondary messenger, cAMP, leading to the reduction of calcium concentration and thus inhibiting platelet activation through granules release as discussed early. PGI2 is used to treat pulmonary arterial hypertension by reducing blood coagulation clotting. Fibrinolysis process reorganizes and reabsorbs mature blood clots by plasmin, a plasma protein synthesized in the liver to inhibit excessive fibrin formation. Plasmin proteolytically cleaves fibrin into fibrin degradation products. Plasmin is regulated by various activators and inhibitors. 

\subsection{Flow effects} 
The formation of hemostatic plug and arterial thrombus starts with the initial adhesion of platelet's glycoprotein Ia/IIa surface receptors to collagen, a main structural protein, exposed at the site of vascular damage. The adhesion process is believed to be critically mediated by red blood cells \cite{p18,p19,p20,p21} which, in turn, depend on rheology. Both shear rate and elongation velocity of the blood flow have an impact on platelet adhesion and deformation. Consequently, the first challenge in the mathematical model of thrombus formation is to accurately model and simulate red-blood-cell-mediated platelet adhesion to collagen.

In addition to the blood flow effect, the initial platelet adhesion in the arterial circulation is strengthened by VWF \cite{p12}. VWF also binds to Factor VIII, which is released from VWF by the action of thrombin. When platelet receptors are activated by thrombin in the amplification process, they also bind to VWF. VWF plays a major role in hemostasis and thrombosis. Bleeding tendency is very common in patients with VWF dysfunction and deficiency \cite{p12,p13}. The latter can be observed in tissues having high blood flow shear in narrow vessels. The structural flexibility of VWF gives rise to its ability to respond with flow motion. Currently, the lack of experimental data makes it difficult to fully understand VWF's spatiotemporal dynamics in blood flow. The modeling of VWF's interaction with blood flow and its binding with collagen and many different platelet receptors is another challenge. Note that clot growth and stability are critically mediated by blood flow field \cite{p6}, while thrombus formation, in turn, has a potentially disturbing effect on blood flow. Intra-thrombus transport processes including obstructed diffusion and advection may also be critical for the spatial control of several stages of thrombosis \cite{p22}. As a result, understanding the interplay between blood flow and hemostasis or thrombosis is one of the main challenges in the field.

\section{Antithrombotic drugs}
Based on the current understanding, a wide variety of anticoagulation drugs has been developed for thrombosis prevention and treatment, and for the risk reduction of ischemic stroke, heart attack, and pulmonary embolism \cite{p23,p24}. For each specific antithrombotic drug or drug candidate, the effect on the coagulation cascade needs to be studied by both animal experiments and clinical trials. So far, most existing antithrombotic drugs suffer from certain degrees of the bleeding side effect.

\begin{figure}[h]
    \centering
    \includegraphics[width=1\textwidth]{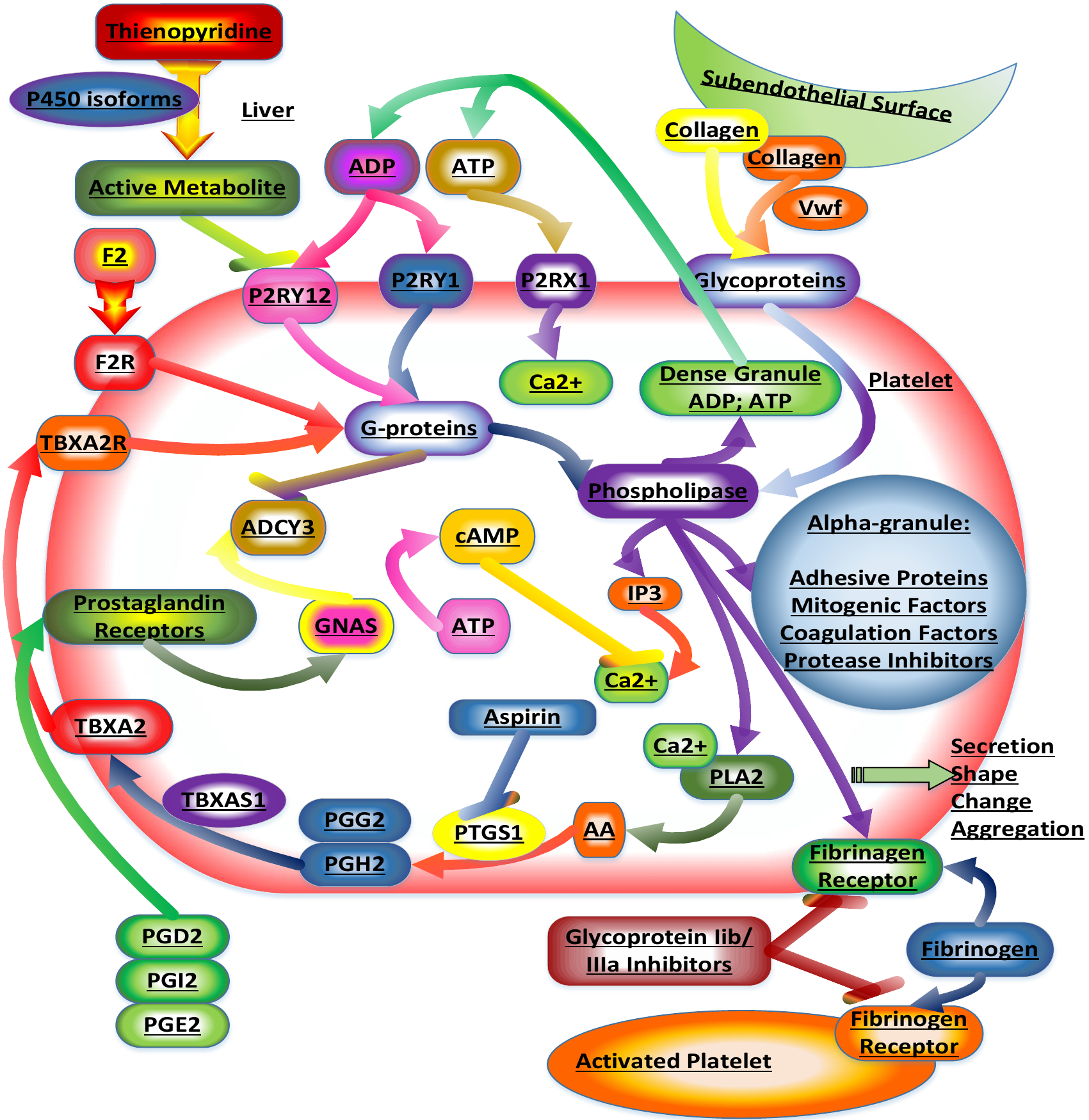}
    \caption{A brief description of platelet activation pathways with an emphasis on drug targets \cite{p24}.  Image credit: Tiffanie Richelle Ma. }
    \label{fig:3}
\end{figure}

\subsection{Platelet activation pathway inhibitors} 

One type of drugs targets the platelet activation pathways as illustrated in Fig. \ref{fig:3}. Among them, Aspirin and Thienopyridine inhibit PTGS1 and P2Y12, respectively, associated with platelet aggregation pathways \cite{p25,p26,p27}.

\subsection{Coagulation cascade inhibitors} 
Most antithrombotic drugs are designed for inhibiting the coagulation cascade of secondary hemostasis. Warfarin (Coumadin) is one of the most commonly used drugs that inhibit the formation and growth of thrombi. As mentioned early, Warfarin inhibits VKORC, an enzyme needed to synthesize vitamin K for supporting mature clotting factors. It inhibits Factors IXa, Xa, VIIa, and prothrombin (II). Another commonly used drug is Heparin, which binds to the enzyme inhibitor antithrombin III. The heparin-activated antithrombin then inhibits thrombin, Factor Xa, and other proteases. Fondaparinux (tiny heparin) binds to antithrombin and inhibits activation of Factor X and the proteolysis of prothrombin (II) as shown in Fig. \ref{fig:2}.  Low-molecular-weight heparin (LMWH) is a class of anticoagulant medications, including Bemiparin, Nadroparin, Reviparin, Enoxaparin, Parnaparin, Certoparin, Dalteparin, and Tinzaparin. Similar to heparin, LMWH drugs inhibit the coagulation process through binding to antithrombin to accelerate its inhibition on activated Factor X (Factor Xa). However, LMWH-activated antithrombin typically does not inhibit thrombin (Factor IIa). Apixaban, Rivaroxaban, and novel oral anticoagulant (NOAC) are Factor Xa inhibitors. Dabigatran, Argatroban, Bivalirudin are thrombin inhibitors. And TFPI inhibits Factor Xa and Factor VIIa, as shown in Fig. 1. Recently, there is a tendency in developing direct Factor XIa inhibitors, which are believed to cause a minor bleeding problem \cite{p28, p29}.

\section{Mathematical Modeling of Thrombosis}
Computer modeling \& simulation can shed light on important aspects of hemostasis and thrombosis \cite{p13,p19,p20}. At present, numerous computational models of physiological and pathophysiological platelet aggregating and clotting have been developed in the past decade. There are no comprehensive models of hemostasis and thrombosis that have been published, but the publications tend to focus on some of the mechanisms in thrombosis, such as hemodynamics, platelet activation/aggregation, and coagulation cascade kinetics. 

Physiological modeling is an approach that applies physiology into computational models by sets of mathematical equations and parameters. Inside the mathematical model, the parameters represent the biochemical, biological and physiological property of the biological system; and the equations represent physiology, pathology, and structure of a biological system. In general, physiological modeling in general mathematically presents only one or two types of physiology and very few are integrative models of more than 3 types of physiology \cite{p30, p31} due to the limited understanding of the interaction mechanisms between different physiological systems.
As a new application of physiological modeling, quantitative systems pharmacology (QSP) modeling has emerged as an approach that integrates biochemistry, systems biology, cellular biology, biomechanics, physiology, and pathophysiology to develop quantitative computational models to solve complex pharmacokinetic and pharmacodynamic problems in drug development \cite{p32}. Some of the fast-growing applications of QSP models are to predict drug dosage and drug efficacy, drug kinetic time profile and drug safety. The scene behind the modeling approach is the underlining mechanisms. To mathematically describe these mechanisms, QSP modeling approaches often include but not limited to systems biology, protein networks, signal transduction pathways, cellular mechanics, biomechanics, systems physiological modeling, clinical data, and virtual patient simulation, and even deep learning \cite{p33,p34}.

\subsection{Coagulation cascade} 

Modeling the coagulation cascade is more straightforward as these processes can be conducted and measured in-vitro.  When blood is removed and placed in a glass test tube, it clots fairly quickly, but the addition of EDTA to bind calcium prevents this from happening.  Clotting can be initiated in vitro at a later time by adding initiating agents such as kaolin or TF, which probe the intrinsic and extrinsic pathways, respectively.  This ability to conduct coagulation experiments in-vitro, allows measurement of the kinetics of the various coagulation factors.  Numerous mathematical models have been developed based exclusively on the kinetics of in-vitro measured coagulation.  One of the earliest models described the generation of thrombin in-vitro via initiation by TF, and included 18 differential equations and 20 reaction rate constants \cite{p35}.  This model was later upgraded to include TFPI and antithrombin III, and has been thoroughly verified in experimental studies \cite{p36,p37,p38}.  This model has been termed the ``Hockin-Mann Model'' and forms the basis around which most of the later coagulation/thrombosis models have been developed.  Burghaus et al expanded on the Hockin-Mann Model by the inclusion of contact/intrinsic pathway activation via Factor XII, and used the model to predict the in-vivo efficacy and safety of different doses of rivaroxaban \cite{p39, p40}.  A similar but independently developed coagulation model was reported by Nayak et al and used to characterize coagulation in normal human plasma vs. Factor VIII deficient plasma, in addition to the effects of the addition of various zymogens \cite{p41}.  This model was later used to predict in-vivo coagulation biomarker changes in patients with hemophilia A or B \cite{p42}. 

\subsection{Incorporation of platelet effects} 
 The coagulation cascade works synergistically with platelets to form a thrombus.  Therefore it is important to include the effects of platelets and the coagulation cascade to characterize and predict thrombus formation.  Modeling of platelet biology was investigated even before models of the coagulation cascade were available.  Simple models of platelet aggregation/disaggregation mediated via ADP were developed in the 1980s \cite{p43,p44,p45}.  However, these early models were focused primarily on simple platelet biophysics.  Moll and Fogelson later incorporated the system of reaction-diffusion equations which describe the chemically-mediated spread of platelet activation and aggregation to better capture in-vivo platelet physiology \cite{p46}.  Luan et al demonstrated how the incorporation of platelet activation/aggregation, as well as the coagulation cascade, allowed identification of the best sources for therapeutic intervention in the prevention of thrombosis in patients \cite{p47}.  Chatterjee et al published a different model incorporating platelet activation and building on the Hockin-Mann coagulation model \cite{p48}.  This model was used to investigate the effects of blockage of Factor XIIa via corn trypsin inhibitor (CTI), thus probing the specific dynamics of that component of the intrinsic pathway.  

\subsection{Incorporation of blood vessel fluid dynamics} 

To better reproduce the physiological conditions of in-vivo thrombus formation, in addition to the inclusion of platelet and coagulation effects, incorporation of the fluid dynamics within the blood vessel is critical.  More recent mechanistic modeling efforts have focused on this challenge, with the help of experimental data from in-vitro microfluidic devices \cite{p49, p50}.  Leiderman and Fogelson, have pioneered the efforts to develop a more comprehensive model of thrombus formation, incorporating all of the spatial–temporal aspects of platelet aggregation and blood coagulation under flow that including coagulation biochemistry, chemical activation and deposition of blood platelets, and the interaction between the fluid dynamics and the growing thrombus \cite{p51}.  This model has been used to identify important and clinically relevant behaviors of thrombosis, such as the stoichiometric dependence of Factor XI activation and platelet counts in the blood vessel \cite{p52}; and the role of fluid and platelet transport on thrombus formation \cite{p9}.  A global sensitivity analysis of the latest version of this model identified a number of variables that were most important in determining the degree of thrombus formation, including platelet count, platelet adhesion rate, the rate of Factor X activation, the rate of Factor V activation by Factor Xa, prothrombin activation, and blood flow rate \cite{p53}.  The results of the sensitivity analysis were then used to evaluate the potential role of Factor V in hemophilia A (Factor VIII deficiency), demonstrating the counterintuitive result that lower Factor V levels and/or activity would result in greater thrombin generation in these patients \cite{p54}.  

\subsection{Utility in drug discovery and development} 
\begin{figure}[h]
    \centering
    \includegraphics[width=1.\textwidth]{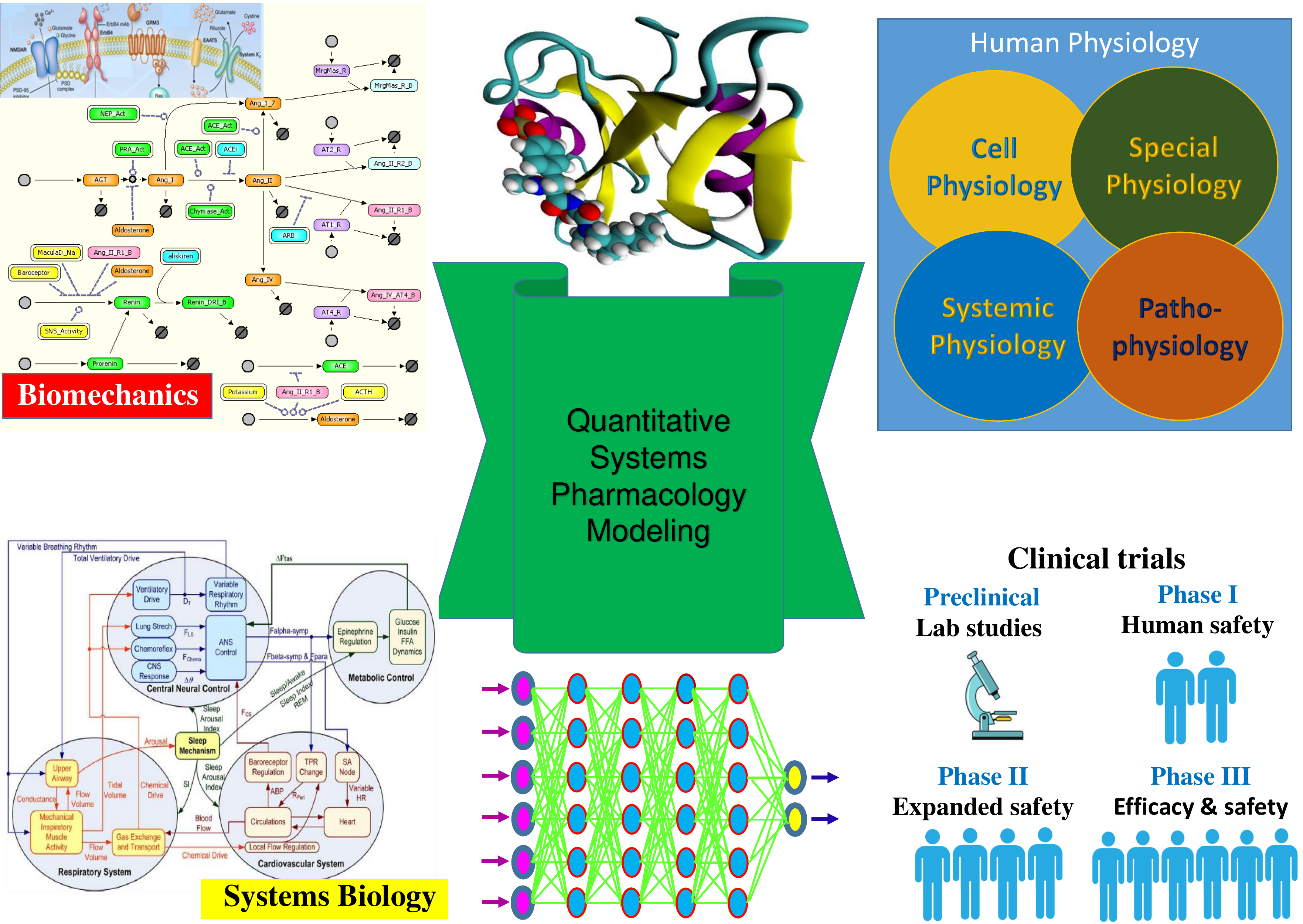}
    \caption{Illustration of quantitative systems pharmacology modeling for drug development, involving biochemistry, systems biology, biomechanics, physiology, patient data, and deep learning.  }
    \label{fig:4}
\end{figure}

One of the most important applications of mechanistic modeling of disease pathophysiology and drug pharmacology (i.e. QSP) is in aiding in the process of hypothesis generation and testing in drug discovery and clinical development (Figure \ref{fig:4}) \cite{p32,p55,p56}.  There are numerous published examples of the use of mechanistic models of thrombosis being used to aid in drug discovery and development.  As mentioned above, a mechanistic model of the coagulation cascade was used to predict the dose-response of rivaroxaban for VTE and bleeding in orthopedic surgery patients; as well as to estimate the optimal timing of switching from warfarin to rivaroxaban \cite{p39,p40,p57}.  These types of predictions can help in the design of clinical trials, and possibly even avoid some clinical trials, making the process of clinical drug development much more efficient.  A simpler model consisting of just the coagulation cascade was shown to accurately predict the pharmacodynamics for warfarin, enoxaparin, heparin, and vitamin K \cite{p58,p59}.  This model was later modified to focus specifically on the pharmacodynamics of Factor Xa inhibitors \cite{p60}.  

\section{Future Directions}
While it is clear that coagulation and thrombosis are one of the most well understood physiological processes, there are still some areas of uncertainty that would benefit from further experimental and computational research.  While the in-vitro coagulation and platelet aggregation are well characterized molecularly and dynamically, the detailed pathophysiology of in-vivo thrombosis requires further exploration.  As discussed previously, there are experimental limitations in measuring the spatiotemporal dynamics, kinetics, and transport of platelet, red blood cells, collagen, and VWF interactions inflow conditions or physiological conditions. Moreover, the problem involves a wide variety of temporal and spatial scales. For example, the blood vessel diameters can vary from 25 mm in the aorta to only 8 µm in the capillaries, making the simultaneous descriptions and simulations of realistic thrombotic kinetic, dynamics and transport impractical at this point.  The interaction of thrombosis with the immune system is another aspect of the pathophysiology of cardiovascular disease that requires in-depth understanding, and would significantly benefit from computational modeling.  Thrombotic models that employ molecular, cellular/tissue, fluid, microfluidic and physiological descriptions have rarely been used in practical QSP for pharmacokinetic and pharmacodynamic computation \cite{p61}. Models of signal transduction \cite{p62, p13, p14} and blood coagulation in vitro \cite{p63, p64} are used to provide limited mechanism-driven molecular-level insights. As a result, simpler models with fewer parameters, rather than complex QSP physiological models, are more often utilized to explore fundamental hypotheses about thrombus generation and drug discovery/development \cite{p6, p10, p11}. Nonetheless, this situation will change gradually as more experimental results become available, which help determine important model parameters of more detailed QSP models.
  
Most existing QSP models are mechanism-based approaches. It is worthy to mention that data-driven models using machine learning, including deep neural networks offer an increasingly important alternative. Machine learning is one of the most transformative technologies in the human history. It has profoundly changed a wide variety of fields of science, engineering, finance, insurance, banking, social media, marketing, retail, etc. \cite{p65,p66,p67,p68}, including healthcare and biomedicine \cite{p69,p70}.  Essentially, mechanism-based QSP approaches exploit chemical, physical, mechanical and physiological laws and/or principles to describe the pharmacokinetics and pharmacodynamics of drugs and diseases in human body. In contrast, data-driven models rely entirely on clinical data to infer possible in vivo drug and disease interactions. A neural network (NN) has been developed to facilitate the patient-specific simulation of thrombosis under hemodynamic and pharmacological conditions \cite{p71}. Using machine learning, including deep learning, to fill the gap between QSP models and available patient data will be a new trend in the field.  

\vspace*{12pt}

\section*{Acknowledgment}
The work of GWW was supported    in part by NIH grant GM126189.


\end{document}